\documentclass[5p]{elsarticle}

\usepackage{lineno}
\modulolinenumbers[5]
\pdfoutput=1
\usepackage{amsmath}
\usepackage{amssymb}
\usepackage{natbib}
\usepackage{graphicx}
\usepackage{subcaption}
\usepackage{txfonts}
\usepackage{booktabs}
\usepackage{siunitx}
\usepackage{longtable}
\usepackage[breaklinks]{hyperref}

\usepackage{cancel}
\usepackage{xcolor}

\usepackage{tabto}

\journal{Astronomy and Computing}

\begin{document}

\begin{frontmatter}

\title{{\it Lenstool-HPC}: A High Performance Computing based mass modelling tool for cluster-scale gravitational lenses}

\author[LastroAddress]{Christoph Sch\"{a}fer\corref{maincorrespondingauthor}}
\cortext[maincorrespondingauthor]{Corresponding author}
\ead{christophernstrerne.schaefer@epfl.ch}

\author[ScitasAddress]{Gilles Fourestey}\author[LastroAddress,LamAddress]{Jean-Paul Kneib}

\address[LastroAddress]{Institute of Physics, Laboratory of Astrophysics, Ecole Polytechnique F\'{e}d\'{e}rale de Lausanne (EPFL), Observatoire de Sauverny,\\ 1290 Versoix, Switzerland}
\address[ScitasAddress]{SCITAS, Ecole Polytechnique F\'{e}d\'{e}rale de Lausanne (EPFL), 1015 Lausanne, Switzerland}
\address[LamAddress]{Aix Marseille Universit\'{e}, CNRS, LAM (Laboratoire d'Astrophysique de Marseille) UMR 7326, 13388, Marseille, France}

\begin{abstract}
With the upcoming generation of telescopes, cluster scale strong gravitational lenses will act as an increasingly relevant probe of cosmology and dark matter. The better resolved data produced by current and future facilities requires faster and more efficient lens modeling software.
   Consequently, we present {\it Lenstool-HPC}, a strong gravitational lens modeling and map generation tool based on High Performance Computing (HPC) techniques and the renowned {\it Lenstool} software. We also showcase the HPC concepts needed for astronomers to increase computation speed through massively parallel execution on supercomputers.
   {\it Lenstool-HPC} was developed using lens modelling algorithms with high amounts of parallelism. Each algorithm was implemented as a highly optimised CPU, GPU and Hybrid CPU-GPU version. The software was deployed and tested on the Piz Daint cluster of the Swiss National Supercomputing Centre (CSCS).    
   {\it Lenstool-HPC} perfectly parallel lens map generation and derivative computation achieves a factor 30 speed-up using only 1 GPUs  compared to {\it Lenstool}. {\it Lenstool-HPC} hybrid Lens-model fit generation tested at Hubble Space Telescope precision is scalable up to 200 CPU-GPU nodes and is faster than {\it Lenstool} using only 4 CPU-GPU nodes.
\end{abstract}


\end{frontmatter}

\section{Introduction}

\begin{figure}
\includegraphics[width=\hsize]{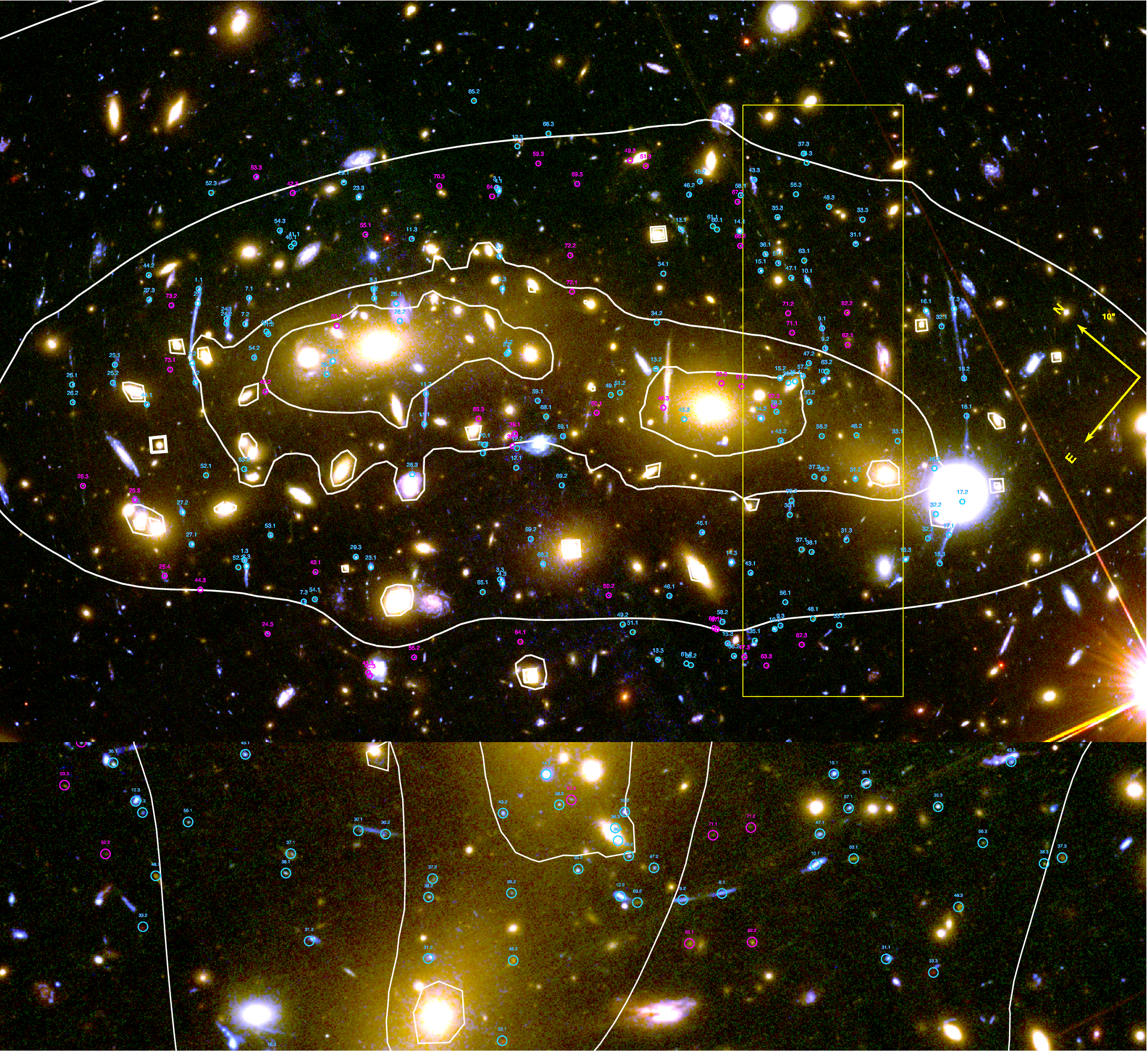}
\caption{MACSJ0416-2403: The cluster has 68 confirmed multiple lensed background sources. The isolines trace the distribution of matter in the cluster which were computed using {\it Lenstool}. The highlighted (green) rectangle represents a zoomframe of the cluster showing the fainter multiple images. Credit. \cite{2014Jauzac}}
\label{fig:MACSJ0416}
\end{figure}

With the advent of high-precision astronomy and big data, high performance computing (HPC) has reached a critical importance for astrophysicists. Astrophysical codes developed over 10 years ago are not able to keep up with the amount of data that new instruments are bringing in. To handle these new challenges it is now necessary to implement HPC techniques and alternative thinking to speed up these softwares. One such example is {\it Lenstool}, a mass modelling tool for strong gravitational lenses. These lenses are rare astrophysical phenomena where a distant light-source is aligned so closely with a foreground galaxy or cluster that its images appears to an Earth observer multiple times. The images appear distorted and magnified similar to objects seen through an unfocused lens. They take the shape of distorted arcs, multiple images and Einstein rings. These distortion are due solely to the gravitational potential of the foreground galaxies or cluster which acts as a lens. This allows specialized mass-modelling software like {\it Lenstool}\footnote{Publicly available at https://projets.lam.fr/projects/lenstool/wiki} \citep{2007Jullo,1996Kneib} to create precise mass-models of the lenses by fitting parametric mass-models\citep{2008Limousin,2011Richard,2015Jauzac} using Bayesian MCMC samplers (see Fig \ref{fig:MACSJ0416} in \cite{2014Jauzac}).

The astrophysical interests are multiple. They are used to study the dark matter profile of lensing galaxies \citep{2015Jauzac} and calculate the dark-baryonic matter ratio \citep{2007Jiang,2011More,2015Sonnenfeld}. Lensed Quasars are used for time-delay studies which constrain the Hubble constant \citep{2016Bonvin,2017Suyu} and the magnification effect of gravitational lenses allows for the study of high-redshift background objects \citep{2004Kneib, 2011Richard, 2015Atek}. 

These precise mass-models are obtained by observers through an iterative process using {\it Lenstool}'s modelling capabilities repeatedly, adding new observational constraints. Using {\it Lenstool} however, especially on deep {\it Hubble} Space Telescope (HST) observations, is becoming extremely time-consuming possibly taking up to one month for one iteration. Beyond slowing down the release of precise mass-models, it severely limits the capability of observers to test new theories for the assembly of mass  in galaxy-clusters. 

To tackle this problem, we developed {\it Lenstool-HPC}, a new parallelism aware library which uses High Performance Computing (HPC) techniques to increase computation-speed by orders of magnitude through parallelisation. {\it Lenstool-HPC} was developed for CPUs and GPUs using CUDA and C++ in a collaboration between HPC experts and astrophysicists. 
The first section presents a brief overview of the theory behind gravitational lensing and {\it Lenstool} mass modelling algorithm and the computational challenge it poses. This is followed by a section summarizing the HPC notions that defined the development of the library before presenting the library itself. We finish this paper with detailed benchmark results of the library, studying in particular the speed-up and scaling of the lensing map generation and mass modeling fit computation compared to {\it Lenstool} on modern CPU and GPU clusters.

\section{Gravitational lens mass-modelling}


\subsection{Gravitational lens theory overview}

A gravitational lens system can to first order be represented by projecting the lens and the source respectively on an image and source plane. 
We usually can assume that the size of a the lens in the line-of-sight direction is negligible compared to the distance between observer, lens and source, this is the "thin-lens" approximation. Then the gravitational lensing phenomenum can be summarized by a simple trigonometric equation called the lens-equation:


\begin{equation}
\vec{\beta}=\vec{\theta} -\vec{\alpha}(\vec{\theta}) \quad,
\label{equ:lensequation}
\end{equation}

where $\vec{\beta}$ is the angular position of the source in the source-plane and $\vec{\theta}$ the angular position of the image in the lens-plane. The deflection angle $\vec{\alpha}$ is the gradient of the lensing potential:

\begin{equation}
\psi(\vec{\theta}) = \dfrac{1}{\pi}\int_{\mathbb{R}^2}d^2\theta'\kappa(\vec{\theta}) \rm{ln}\vert \vec{\theta}-\vec{\theta}'\vert \quad,
\end{equation}

where $\kappa(\vec{\theta}) $ is the surface mass density of the lens-plane defined as 

\begin{equation}
\kappa(\vec{\theta}) = \dfrac{\Sigma(D_d\vec{\theta})}{\Sigma_{crit}} \qquad \rm{with} \quad \Sigma_{crit}= \dfrac{c^2}{4\pi G}\dfrac{D_s}{D_l D_{ls}} \quad,
\end{equation}

and $\Sigma_{crit}$ is the critical surface mass density. $D_s$, $D_l$ and $D_{ls}$ are respectively the distance from the observer to the source, to the lens and between lens and source. The lensing potential $\psi(\vec{\theta})$ is the normalised Newtonian gravitational potential, satisfying the relations $\vec{\alpha} = \nabla \psi$ and $\kappa = \nabla^2 \psi$. 

The  distortion of the images described by the following Jacobian matrix (the magnification matrix) is derived from the lens equation:

\begin{equation}
\vec{A}^{-1}(\vec{\theta}) = \dfrac{\partial \vec{\beta}}{\partial \vec{\theta}} = (\delta_{ij} - \dfrac{\partial^2\psi(\vec{\theta})}{\partial\theta_i\partial\theta_j}) =
\begin{pmatrix}
1 - \kappa -\gamma_1 	& - \gamma_2				\\ 
- \gamma_2 				& 1 - \kappa +\gamma_1   	\\
\end{pmatrix}\quad,
\end{equation}

where $\gamma_1 $ and $\gamma_2$ are the shear components, quantifying the amount and direction of the gravitational shear.

The magnification value is related to the determinant of the Jacobian matrix as 

\begin{equation}
\mu(\vec{\theta_0}) = \dfrac{1}{\mathrm{det}(\vec{A}^{-1})}.
\end{equation}

\begin{figure}
 \includegraphics[width=\hsize]{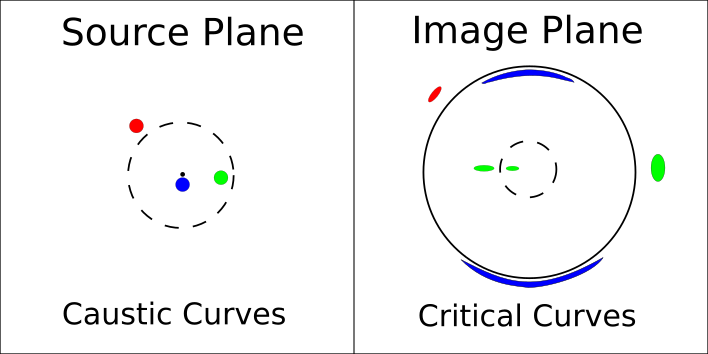}
 \caption{Schematic of the lensed images formed by three sources due to the gravitation potential of a non singular isothermal sphere. The red source is outside the caustic lines therfore has only lensed image. The green source is inside a caustic line and is lensed three times. The blue source is almost perfectly aligned with the center of the lens and is starting to form an Einstein ring.}
 \label{fig:lens_system}
 \end{figure}

The points where $\mathrm{det}(\vec{A}^{-1}) = 0 $ form the critical lines where the magnification is theoretically infinite. In practice the wave-nature of light leads to finite amplification. Their unlensed counter-part in the source plane are called caustics. These caustics set the boundaries of areas where the image of a source is not just distorted but also multiplied. Every source which moves across will have two more or less lensed images (fig. \ref{fig:lens_system}). More details on lensing theory can be found in \cite{2001Bartelmann}.

\subsection{Mass-modelling}

{\it Lenstool} \citep{1996Kneib,2007Jullo,2009Jullo} creates mass-models for lensing cluster by fitting parametric mass-models to each individual cluster sub-halos. 
Depending on the parametric model used, the free parameters can vary. They include generally the position of the center of the sub-halo, its dispersion velocity, its ellipticity and orientation, as well as other free parameters specific to the model \citep{2007Elliasdotir} in particular related to the mass profile. 
In clusters of galaxies, the main constraints used for fitting are the position of identified multiple lensed images. 

\begin{figure}
\includegraphics[width=\hsize]{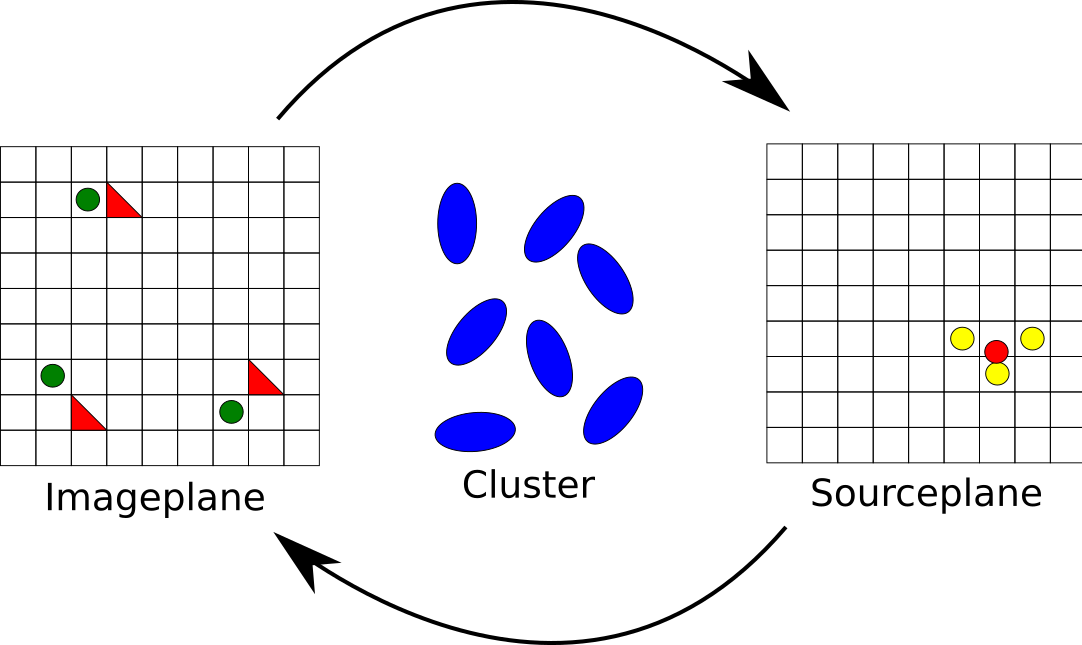}
\caption{{\it Lenstool} Mass modelling: Multiply imaged sources (green dots) work as constraints. Each image position is lensed onto the source-plane (yellow dots) using the current mass-model. The barycentre of these constraint (red dot) is taken as the best approximation of the source position and then lensed back into the lens-plane (red triangles). The difference between the constraints and lensed back source approximation gives an approximation of the fit of the mass-model.}
\label{fig:lenstool_chi}
\end{figure}

Each image is unlensed onto the source-plane using the mass-model to be tested. In the case of a perfect model all images should end up at the same point in a source plane. However, in practice the model is off, so the corresponding sources of the multiple-images are at slightly different positions. Sending back the barycentre of these positions to the image plane (see Fig. \ref{fig:lenstool_chi}), we can define a cost-function that {\it Lenstool} will try to minimize:

\begin{equation}
\chi^2 = \sum^{N}_{i} \chi_{i}^2 = \sum^N_i \sum^{M_i}_j  \dfrac{(c_{ij}-x_{ij})^2}{\sigma^2_{ij}}
\end{equation}

where $N$ is the number of lensed sources, $M_i$ the multiplicity of those sources, $c_{ij}$ the multiple-image constraints, $x_{ij}$ the back and forth lensed constraints and $\sigma^2_{ij}$ the error-budget. The exploration of the parameter space and of the optimum solution is done using an Bayesian Markov Chain Monte Carlo Algorithm (MCMC). More details on the procedure can be found in \cite{2007Jullo,1996Kneib}.

The number of optimized free parameters depends on the parametric model used but can range up to a thousand for a typical cluster-lens model (see \cite{2014Jauzac,2015Jauzac}). In the high dimensionality of the problem lies the first computational challenge from {\it Lenstool}.  
Even using an Bayesian MCMC algorithm, {\it Lenstool} has to try an enormous amount of parameter-combinations to find solutions that minimize the cost-function. 




\subsection{$Chi^2$ computation}

\begin{figure}
\includegraphics[width=\hsize]{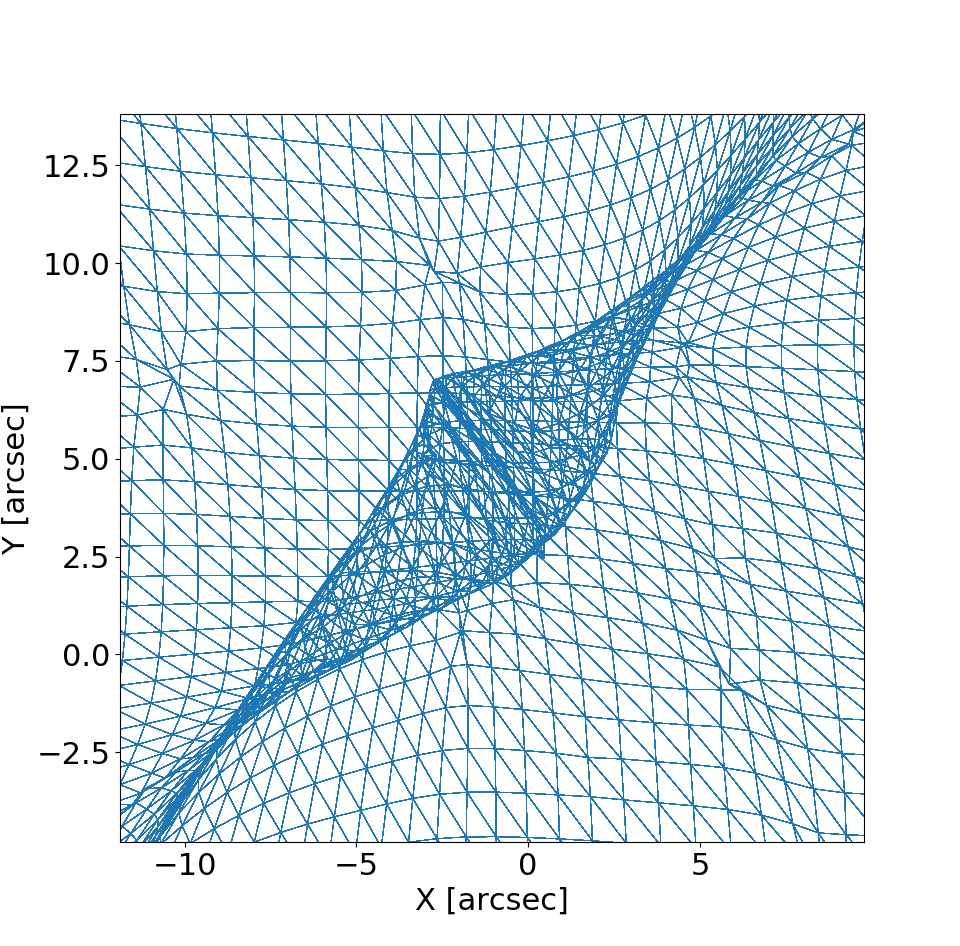}
\caption{Graphical representation of the unlensing of the quadratic triangular grid from the image-plane unto the source plane. The lines which delimit the area where the grid folds unto itself (where therefore multiple images can be found) are the caustic lines. 
}
\label{fig:GridSource}
\end{figure}

The second computational challenge is the $Chi^2$ computation based on the unlensing and relensing of multiple imaged sources. Unlensing a point into the source-plane is a simple but non revertible application of the lens-equation (equation \ref{equ:lensequation}). The multiple solutions for the relensing problem can as a consequence not be computed analytically. To compute predicted multiple images of a source, the "brute force" approach is to unlens a image-plane grid unto the source-plane and check each quadrant for the presence of the source (see Fig.\ref{fig:GridSource}).  {\it Lenstool} avoids this computationally costly approach by using a variant of the "image-transport" method \cite{1992Schneider}.

The method works as follows: It defines a triangle around the constraint that does not contain any other constraint but is likely to contain the source. The triangle is then subdivided into 4 smaller triangles. Each triangle is checked for the source. If a triangle containing the source is not found, the immediate environment of the triangle is searched. The subdividing process is continued recursively until a precision of $10^{-4}$ arcsec is reached. While a lot faster than the brute force approach, this method is not fully stable. Extremely strong amplification near the critical lines can degrade the zoom-in process sufficiently to lose images which can complicate the modelling process. 

\subsection{Lensing Maps}

The third computational challenge we address is the computation of lensing maps. Lensing maps are used to visualize crucial information of the lens-system. Each map is organized into a rectangular grid defined on the image-plane, each grid cell usually being the size of a pixel of image data. Information that can be visualised are the projected surface mass density $\kappa$ of the lenses, the projected shear $\gamma$ (norm, direction, individual components), the amplification $\mu$ or its inverse, the lens deflection field, the lensing potential $\varphi$, the time delay surface  
and variations thereof. 
More information on these maps can be found at: \url{https://projets.lam.fr/projects/lenstool/wiki}.

Lensing-maps are also used to calculate the statistical error inherent to the Bayesian process. In order to compute the map variances, full-resolution lensing maps have to be generated for each tested parameter-combination which is a time-consuming task. 

The critical part of the lensing map computation is the second order derivatives of the gravitational potentials of each cluster member, which allow to compute $\kappa$, $\gamma$ and $\mu$:



\begin{equation}
\kappa(x,y) = \sum^{N_h}_i \dfrac{1}{2}\left( \partial_{xx}\phi(x,y) + \partial_{yy}\phi(x,y) \right)
\label{equa:Map1}
\end{equation}


\begin{equation}
\gamma^2(x,y) = \sum^{N_h}_i \dfrac{1}{4}\left(\left(\partial_{xx}\phi(x,y) - \partial_{yy}\phi(x,y) \right)^2 + \left(\partial_{xy}\phi(x,y)\right)^2\right)
\label{equa:Map1}
\end{equation}

\begin{equation}
\mu = ((1 - \kappa)^2 + \gamma^2 )^{-1} 
\label{equa:Map1}
\end{equation}

with $N_h$ the number of halos, which includes the large scale components and the sub-halos (attached to each cluster galaxy). At each grid-point the second-order derivative contribution of each cluster member is added up to compute the total derivative. The advantage of {\it Lenstool}'s parametric mass-models is that their single and double derivative can be explicitly calculated through analytical function rather that through a numerical calculation. This makes the computation of the various lensing properties fast, as analytically calculated gradient are faster to compute and do not suffer the numerical errors introduced by numerical derivation and interpolation.


Despite this advantage, the computational challenge is impressive. For Abell 2744 \cite{2015Jauzac} error calculation (one of the Hubble Frontier Field Cluster [HFF]), 10014 maps with 6000x6000 pixels had to be generated. With 258 parametric potentials this corresponds to $10^9$ derivatives per map for a grand total of $10^{14}$ derivative computations. The total process adds up to a total of 300 CPU hours using {\it Lenstool} just for the map generation. 


\section{High Performance Computing (HPC)}

Due to the impossibility of increasing much further the clock-frequency of processors \citep{2001Mudge}, hardware development focus has gone into integrating multiple cores capable of multiple simultaneous operations. This was motivated by Little's Law \citep{1997Bailey}, which states that the performance of computation can be increased through parallel execution. In other words performance can be improved by distributing the work on multiple computation units. Multicore CPUs and GPUs are the consequences of this design choice. This increasingly parallel execution orientated development does not work well with parallelism unaware (often serial) algorithms like {\it Lenstool}'s Image transport method which lack the necessary concurrency for parallel execution. This creates large performance gaps called the Ninja Gap \citep{2012Satish}. 

The following chapter introduces a few essential concepts of High Performance Computing (HPC) necessary to understand how to remove this gap: 1) how performance for software is defined and can be improved and 2) how to implement the different parallelism strategies on CPU and GPUs.

\subsection{Software Performance and Parallelism}

The performance of a software, better known as its throughput, is defined as the number of Floating-point operation per second [flop/s] it is capable of performing. Little's Law states that the throughput ([flop/s]) of a computation is equal to the level of parallel computation instances divided by the latency ([s]). Latency is defined as the time of a single computation instance to process and store an operation. The amount of parallelism that a software can reach is directly related to the level of concurrency the underlying algorithm possesses where concurrency refers to the ability of an algorithm to execute parts of itself out of order without affecting the final outcome.

\begin{center}
\begin{equation*}
\rm{Throughput} = \dfrac{\rm{Parallelism}}{\rm{Latency}}
\end{equation*}
\end{center}

The obvious consequence of Little's Law is that it is possible for software with high parallelism but also higher latency to achieve a higher performance than non parallel low latency software. To achieve optimum computation speed it is therefore necessary to choose carefully the underlining algorithms so as to be "parallelism aware" meaning balancing a high level of concurrency with low latency.

Increasing performance can therefore either be done by reducing latency or increasing concurrency to fully use the available parallel computation capabilities of the hardware. HPC tends to focus on the latter. 

\subsection{Hardware}

Software computation speed is extremely dependent on the hardware it runs on. CPUs and GPU rely on different parallelism strategies to achieve an optimum throughput which need to be taken into account in the development. Multicore CPUs are mainly designed for single thread performance \citep{2010Holzle}. Their lower latencies makes them ideal for less parallelisable applications that use irregular patterns or data structures. GPUs in contrast are designed for massively parallel software. Their individual threads are slow but the much higher number of them allows to hide their high latency and achieve a high throughput on problems with a high number of simple and parallelisable computation.

\subsubsection{Parallelism on CPUs}

  \begin{figure}
  \includegraphics[width=\hsize]{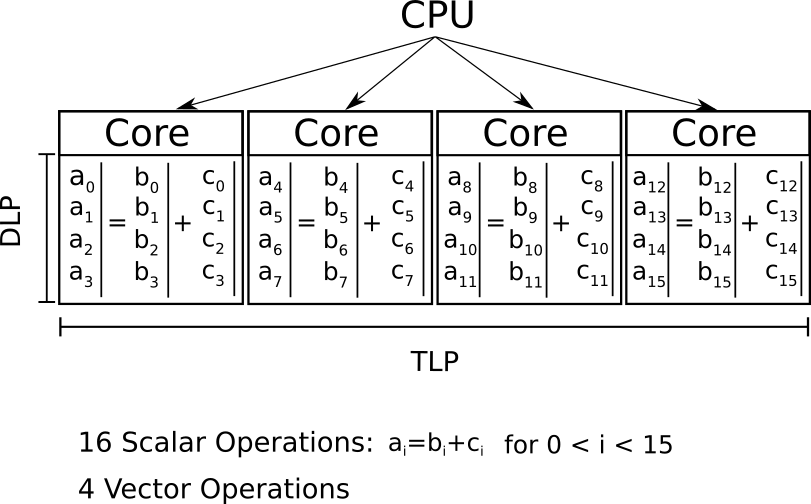}
  \caption{Scaling at a node level: This example CPU is AVX capable, meaning each of his cores is capable computing 4 scalar operations in parallel. This CPU can compute 16 scalar operations distributed over 4 cores (TLP) in vectors of size 4 (DLP) simultaneously. 
  }
  \label{fig:CPUlevels}
  \end{figure}

A CPU consists of multiple cores sharing memory, each capable of executing different independent tasks. Each core can also execute multiple operations simultaneously for the same task by generalizing scalar operations to vectors and matrix operations. At a single CPU (node) level, parallelism is typically divided into three levels: Thread-level parallelism (TLP), Data-level parallelism (DLP) and Instruction-level parallelism (ILP).

TLP optimizes the concurrent execution of tasks (threads) between the different cores, handled by libraries such as OpenMP, Intel's TBB or POSIX pthreads. It mainly handles the problems that come from sharing resources like the memory.

\begin{figure}
  \includegraphics[width=\hsize]{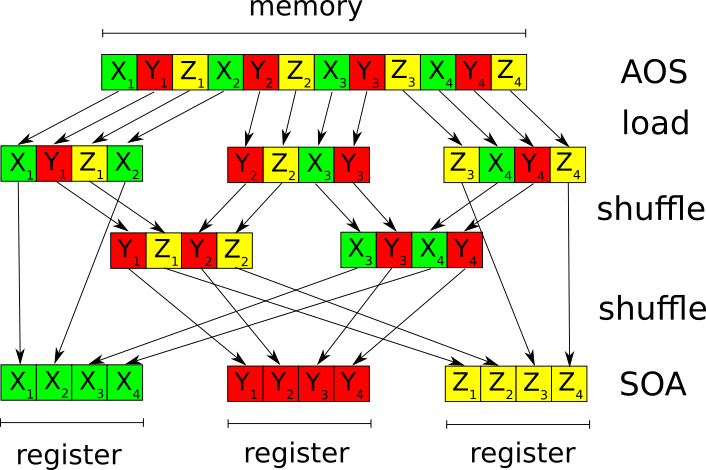}
  \caption{Preparation for vectorisation with a heterogeneous memory and AOS structures: The CPU core first loads from the main memory the needed information into AVX registers. Those registers have to be shuffled multiple times to achieve the needed homogeneous layout. Once the computations are done, they have to be reshuffled back into the AOS structure. Beyond the obvious time loss, the compiler is not able to vectorise these operations automatically. If developers still wish to implement AOS structures, SIMD pragmas have to be used to vectorise the operations manually. $X_i$, $Y_i$ and $Z_i$ represent fictional position information.}
  \label{fig:SOAAOS}
  \end{figure}

DLP handles the vectorisation of scalar operations on a single CPU core using Advanced Vector Extension (AVX). AVX2 and AVX-512 (Advanced Vectorisation Extension) capable CPUs can vectorise respectively 4 to 8 scalar operations through the SIMD (Single Instruction Multiple Data) programming model \citep{2011Kreinin} (Fig.\ref{fig:CPUlevels}). This additional parallelism level comes theoretically at a low development cost. Most compilers are capable of doing implicit vectorisation without developer input by identifying vector operations in the algorithm \citep{2015Intel}. Vector operations however require that the AVX registers are loaded homogeneously with the necessary information using Data structures of type Structure of Array (SOA). Data structures of SOA type stores data of the same type into one parallel array contrary to the more conventional Array of structure (AOS) which interleaves the information(see Fig.\ref{fig:SOAAOS}) \citep{2013Besl,2013Intel}.  


ILP leverage's the superscalar capabilities of modern CPUs, allowing multiple independent instructions to be handled at once. This is mainly handled by the compiler and fall outside of the scope of this paper \citep{2019Intel}.  

\subsubsection{Parallelism on GPUs}




Originally developed for gaming, GPUs are composed of multiple Streaming Multiprocessors (SM) each consisting of multiple Streaming Processors (SP). SP are capable of computing arithmetic operations and are grouped together into warps which share instruction sets. 

The important difference between GPUs and CPUs is that GPUs are not designed for single thread performance \citep{2010Holzle}. GPU threads have much higher latencies than CPUs for floating point operations and memory transfer. To maximise throughput, GPUs are designed to be massively multithreaded. Using the SIMT (Single Instruction, Multiple Threads) programming model, GPUs have hardware threading support that allows hundreds of threads to be active simultaneously, each computing operations in parallel \citep{2008Lindholm,2012Nvidia}. 

The downside of this approach is that if an algorithm has a low level of concurrency, its GPU throughput will be dominated by the high latency \citep{2010Valkov,2013Liang}. If the problem does not propose enough parallel computation to hide the high latency, computation speed will be extremely slow. This makes GPUs compared to CPUs limited in their choice of problems.

Another important aspect of GPU optimisation is paying attention to the ILP problems like divergent execution paths. CPU compilers tend to extract ILP more efficiently than GPUs using modern techniques like Out-Of-Order or speculative execution \citep{2019Intel} without any developer input needed. While DLP is implicitly optimised by the SIMT model\citep{2011Kreinin}, ILP for GPUs has to be explicitly coded.

\section{{\it Lenstool-HPC}}

{\it Lenstool} is comprised of three crucial computations which constitute a bottleneck and can be parallelised: the computation of the deflection potential gradient over a grid, the computation of the $\chi^2$ and the MCMC sampler. {\it Lenstool-HPC} has to date fully optimised the first two of those computations. The gradient computation over a grid is a trivially parallelisable problem with no need for communication between the different parallel tasks for which {\it Lenstool-HPC} proposes an CPU-OpenMP and an GPU based solution. It is vectorisable and has enough parallel computation to hide the GPU latencies. In contrast the computation of the $\chi^2$ is a typical example of a non trivially parallelisable algorithms. It possesses divergent execution paths controlled by the presence of a source in a triangle, atomic operations which cannot be parallelised and imposes a certain amount of communication between the different tasks. For this {\it Lenstool-HPC} proposes a pure-CPU based and a mixed CPU-GPU implementation of the brute-force approach.


\subsection{Gradient Computation}

Computing the various lensing maps or the brute force computation of the $\chi^2$ necessitates the computation of the deflection potential gradient over the whole image. This is done by defining a rectangular grid over the image. For each point the gradient can be calculated analytically from the deflection potentials modeled by multiple parametric potentials. The total gradient in a certain point is simply the sum of the first order derivative of all parametric potentials at that specific point:

\begin{equation}
\nabla \phi(x,y) = \sum^{N_h} \phi_i(x,y)
\end{equation}
where $N_h$ is the number of parametric potentials. 

The gradient computation of different points are independent of each other. Both the CPU-OpenMP and GPU implementation can therefore distribute the task of computing the gradient of a single point to separate computation units. The CPU version uses the implicit vectorisation capability of the intel compiler to additionally vectorize the gradient computation of a single point.

 

\subsection{$\chi^2$ Computation}

  \begin{figure*}
  \includegraphics[width=\hsize]{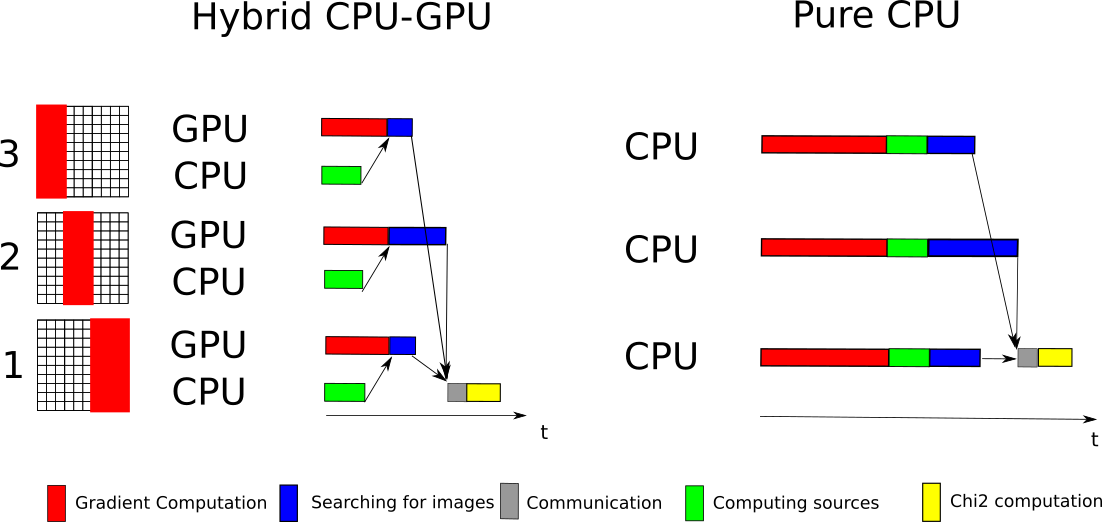}
  \caption{Each GPU is assigned a part of the grid where it computes the gradient. During that time the CPU computes the positions of the sources in the source-plane and sends the information to the GPUs. Once the gradients and the sources are known, the GPUs can start searching for the images by delensing and checking the grid for sources, again by subdividing the grid. Each image found is stored temporarily and at the end of the computation send to the CPU. This operation possesses a divergent execution path, based on if an image is found or not. As a consequence the computation incurs an overhead based on the different amount of found images in each GPUs operational territory. Once all images have been found and received, the master CPU assigns them to their closest constraint and then computes the $\chi^2 $. The purely CPU-based implementation distributes the work similarly to the Hybrid CPU-GPU version with the exception that all CPU cores calculate the same sources positions.}
  \label{fig:HybridGPU-CPU}
  \end{figure*}

In contrast to computing deflection gradients, the image transport method based $\chi^2$ computation is extremely difficult to parallelise. To be able to efficiently distribute the work over multiple computation units, {\it Lenstool-HPC} therefore uses the more computationally intensive but less serial brute-force approach algorithm with GPUs. 
The algorithm can be subdivided into four main stage: The gradient computation of a grid, the source computation based of the constraints, finding images by delensing and checking the grid, and computing the $\chi^2$ based on the found images. 

The distribution of these tasks in {\it Lenstool-HPC} Hybrid CPU-GPU implementation is summarized in Fig. \ref{fig:HybridGPU-CPU}. The gradient computations are divided among the available GPUs. During that time the CPU computes the positions of the sources in the source-plane and sends the information to the GPUs. Once the gradients and the sources are known, the GPUs can start searching for the images by delensing and checking the grid for sources, again by subdividing the grid. Each image found is stored temporarily and at the end of the computation send to the CPU. This operation possesses a divergent execution path, based on if an image is found or not. As a consequence the computation incurs an overhead based on the different amount of found images in each GPUs operational territory. Once all images have been found and received, the master CPU assigns them to their closest constraint and then computes the $\chi^2 $.

The purely CPU-based implementation distributes the work similarly to the Hybrid CPU-GPU version with the exception that all CPU cores calculate the sources positions.



\subsection{Implementation}

We developed {\it Lenstool-HPC} to be similar to {\it Lenstool} to assure continuity for {\it Lenstool} users. {\it Lenstool-HPC} can be compiled as a library with the above mentioned functions and as an executable with the same image-plane mapping capabilities as {\it Lenstool}. All mapping methods have been tested against the corresponding {\it Lenstool}-maps and found correct inside the boundaries of numerical errors. The $\chi^2$ computation is for the moment only available as a function of the library for future MCMC development. The executable works in exactly the same way as {\it Lenstool}, with a master parameter file, and separate file for constraints and mass-modelling potentials as described in the {\it Lenstool} wiki \footnote{\url{https://projets.lam.fr/projects/lenstool/wiki}}. The $\chi^2$ computation is also resistant to missing image problem near caustic lines because of the brute-force approach used. The software and installation instruction can be found at \url{https://git-cral.univ-lyon1.fr/lenstool/LENSTOOL-HPC}.

\section{Results and benchmarks}

This result and benchmark section is organized as follows: First an analysis of the effects of CPU vectorisation and GPUs on the gradient computation. Second a study on the scaling of the CPU and GPU implementation of the $\chi^2$ computation. The scaling studied here is the strong scaling, meaning the same amount of operations distributed over more Nodes. 

The Benchmark configurations were taken from an example strong lensing model of MACS\,J1149.5+2223 from here on named M1149. It is made of 217 different potentials, modelling the cluster. To constrain the model 80 sources have been generated, adding to a total of 227 multiple images. The grid spans over 150 by 150 arcseconds and has 5000 by 5000 pixels for a typical Hubble sampling of 0.03 arcseconds (resolution of $\sim$0.1 arcsec or better).
The Benchmarks were run on five different clusters summarized in table \ref{table:Hardware}. We have chosen to concentrate on the Helvetios CPU cluster and the Piz Daint hybrid CPU-GPU cluster. Both are comprised of the most modern CPU and GPUs on the market we had access to at the writing of this paper. This will allow us to compare the peak performance of the CPU and GPU version of {\it Lenstool-HPC}. To enable a fair comparison of the single-map generation algorithm, we upgraded {\it Lenstool}'s algorithm to support multicore parallelism, distributing the computational operation in the same way as {\it Lenstool-HPC}'s CPU version over the multiple cores. {\it Lenstool} has already OpenMP parrallelisation in its native code but it is only implemented in its multi-map generation algorithm.

\begin{table*}
\caption{Characteristics and sustained performance of Computing Cluster used for the {\it Lenstool-HPC} benchmarks. (*DDR4/MCDRAM)}             
\label{table:Hardware}  
\begin{tabular}{r|lllll}
\hline
 \mbox{name} & \mbox{Piz Daint CPU} & \mbox{Piz Daint GPU} & \mbox{Tave} & \mbox{Helvetios} & \mbox{Fidis} \\
\hline
\hline
\mbox{CPU type} & E5-2695 v4 & E5-2690 v3 & Xeon Phi 7230 & Xeon Gold 6140 & E5-2690 v4\\
\mbox{Microarchitecture} & Broadwell & \mbox{Haswell} & \mbox{Knight's Landing} & \mbox{Skylake} & \mbox{Broadwell} \\
\mbox{Number of cores} & 36 & 12 & 64 & 36 & 24 \\
\mbox{Frequency (Ghz)} & 2.1 & 2.6 & 1.3 & 2.3 & 2.6 \\
\mbox{Memory size (GB)} & 64 & 64 & 112/16* & 192 & 128\\
\hline
\mbox{FP Peak (Gflops/s)} & 1200 & 488 & 1785 & 2136 & 1068 \\
\mbox{stream copy (GB/s)} & 116 & 59.7  & 87/465* & 164 & 120 \\
\hline
\hline
\mbox{GPU type} &  & \mbox{P100} &  &  &  \\
\mbox{Frequency (Ghz)} & & 1.126 & & &   \\
\mbox{Memory size (GB)} & & 16 & & &   \\
\hline
\mbox{FP Peak (Gflops/s)} & & 4546 &  & &  \\
\mbox{stream copy (GB/s)} & & 489 &  & &  \
\end{tabular}
\end{table*}

\subsection{Core Scaling analysis}

  \begin{figure*}
    \begin{subfigure}[b]{0.55\textwidth}
        \includegraphics[width=\textwidth]{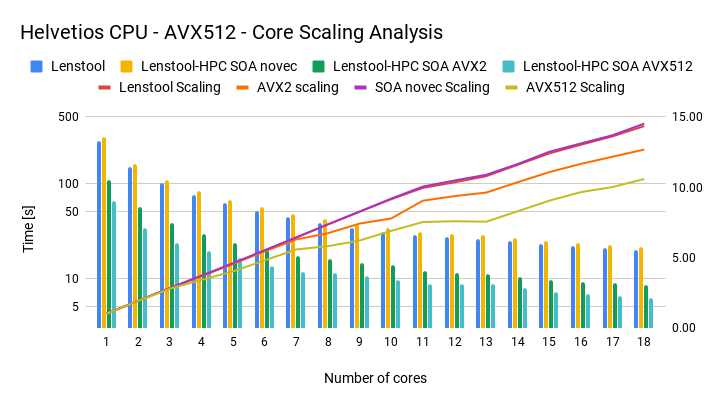}
        \caption{Core Scaling analysis results on AVX512 capable Helvetios cluster. }
        \label{fig:Piz_Daint_GPU_Gradient}
    \end{subfigure}
    \begin{subfigure}[b]{0.45\textwidth}
        \includegraphics[width=\textwidth]{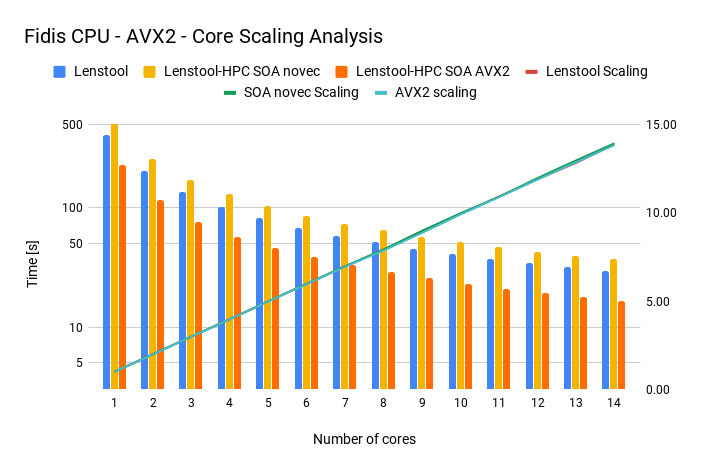}
        \caption{Core Scaling analysis results on AVX2 capable Fidis cluster.}
        \label{fig:Helvetios_CPU_Gradient}
    \end{subfigure}
      \caption{Core Scaling analysis results: In histograms are compared {\it Lenstool}, {\it Lenstool-HPC} with SOA layout without vectorisation (novec) and with vectorisation (SIMD). We observe a speedup of factor 4 for AVX512 machines and factor 2 for AVX2 machines compared to {\it Lenstool}. Without vectorisation {\it Lenstool-HPC} with SOA layout is slightly slower then {\it Lenstool}, indicating that for gradient computation AOS layouts allow for faster memory access then the SOA layout. {\it Lenstool-HPC} AVX scaling diminishing in function of cores (orange and green line) on Helvetios also indicates that memory overheads are getting significant and that we are hitting hardware limits. }
  \label{fig:Benchmark_CoreScaling}
  \end{figure*}


First we studied the scaling at a single processor level, meaning how well it scaled on multiple cores. The benchmark task was to compute one full 5000x5000 gradient map for the M1149 model. Compared were {\it Lenstool}, {\it Lenstool-HPC} using AOS structures, {\it Lenstool-HPC} using SOA structures and no vectorisation and {\it Lenstool-HPC} using SOA structures with vectorisation. The results are summarized in Fig. \ref{fig:Benchmark_CoreScaling} and are detailed in table \ref{table:Helvetios_Core} and \ref{table:FIDIS_Core} . They were run ten times each on Helvetios with AVX 512 capable machines and on Fidis with AVX2 capable machines and no significant standard deviation was observed. An additional Benchmark was run on Helvetios with the amount of zmm-registers limited to AVX2 levels.

It is immediately obvious that on Helvetios, {\it Lenstool-HPC} with vectorisation is indeed faster than {\it Lenstool} by approximately a factor 4. This does not correspond to our theoretical expectations for AVX 512 capable machines which use vectors of size 8 for double-precision floating point operations. This almost twice slower behaviour seems to be due to Intel limiting the frequency of the cores depending on the workload. According to \cite{2017Intel,2019Intel}, the AVX512 and AVX2 top frequency is limited at a lower rate then the non AVX one because of the differing thermal and electrical requirements. The results on the slower AVX-2 capable Fidis machine and the AVX2-limited Helvetios run seem to confirm this. They show the same tendencies as on Helvetios with a speed-up gained by vectorisation around 1.77 for Fidis and 2.22 for AVX2 limited Helvetios which corresponds roughly to half the theoretically expected factor 4. 

It is interesting that, when deactivating vectorisation with the compiler flag no-vec, {\it Lenstool} actually performs better then the {\it Lenstool-HPC} SOA version. For comparison purposes, we created a {\it Lenstool-HPC} AOS version with the results shown in table \ref{table:Helvetios_Core} and table \ref{table:FIDIS_Core} which improves on the {\it Lenstool} results. The speed-up due to vectorisation is however still significant enough to beat our own AOS version. This lower performance by the non vectorised SOA version could suggest that the memory access using SOA layout is not optimised for the gradient computation but more in detailed tests would be neccesary to be certain. 

In the Helvetios results, we also observe a decrease in parallelism efficiency the more cores are used. This is probably due to bandwidth saturation \citep{2010Intel} because of the increased amount of information used by AVX operations. AVX512 operations use 8 times more information than non vectorised operations and 2 times more then AVX2. The decrease in efficiency over 18 cores is not too important but it does show that we are starting to approach the hardware limits of actual CPUs. FIDIS does not show the same trend because even with a 4 times increase in speed due to AVX2, bandwith saturation will not be significant compared to the total operation time.


\subsection{Distributed Scaling}

  \begin{figure*}[h]
    \begin{subfigure}[b]{0.5\textwidth}
        \includegraphics[width=\textwidth]{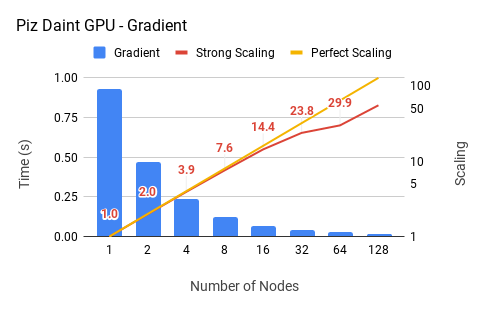}
        \caption{Benchmark results for the deflection gradient computation using the hybrid GPU-CPU version.}
        \label{fig:Piz_Daint_GPU_Gradient_3}
    \end{subfigure}
    \begin{subfigure}[b]{0.5\textwidth}
        \includegraphics[width=\textwidth]{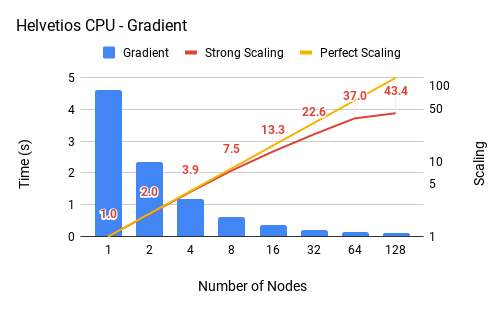}
        \caption{Benchmark results for the deflection gradient computation using the CPU version.}
        \label{fig:Helvetios_CPU_Gradient}
    \end{subfigure}
    \begin{subfigure}[b]{0.5\textwidth}
        \includegraphics[width=\textwidth]{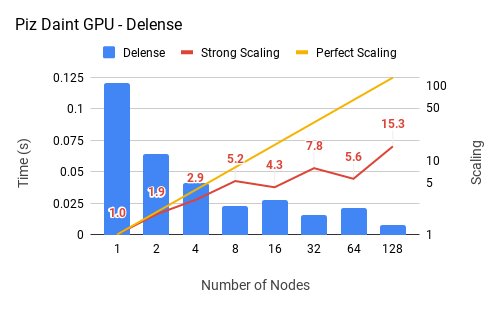}
        \caption{Benchmark results for the constraint delensing operation using the hybrid GPU-CPU version.}
        \label{fig:Piz_Daint_GPU_Delense}
    \end{subfigure}
    \begin{subfigure}[b]{0.5\textwidth}
        \includegraphics[width=\textwidth]{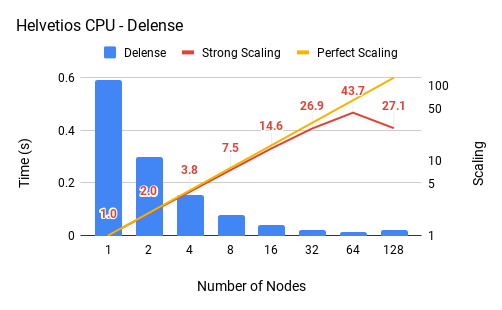}
        \caption{Benchmark results for the constraint delensing operation using the CPU version.}
        \label{fig:Helvetios_CPU_Delense}
    \end{subfigure}
    \begin{subfigure}[b]{0.5\textwidth}
        \includegraphics[width=\textwidth]{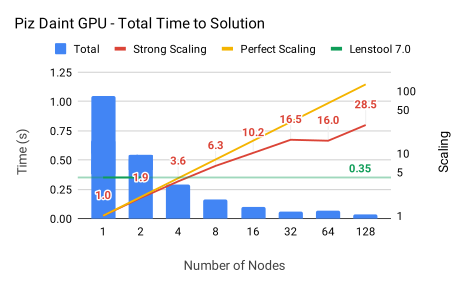}
        \caption{Benchmark results for the total computation time using the hybrid GPU-CPU version.}
        \label{fig:Piz_Daint_GPU_Total}
    \end{subfigure}
    \begin{subfigure}[b]{0.5\textwidth}
        \includegraphics[width=\textwidth]{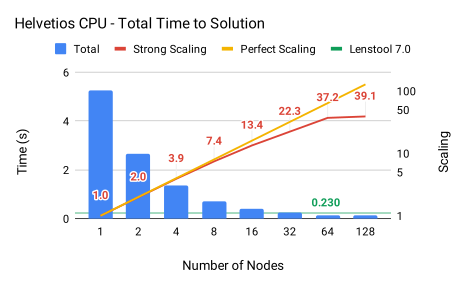}
        \caption{Benchmark results for the total computation time using the CPU version.}
        \label{fig:Helvetios_CPU_Total}
    \end{subfigure}
  \caption{Benchmark and Scaling results of {\it Lenstool-HPC} on Pizdaint GPU (CSCS) and Helvetios(EPFL): the blue histogram shows the time results in function of the number of computation units (nodes) used. The yellow and red line indicate respectively the ideal and actual scaling of Lenstol-HPC computation time in function of nodes used. The horizontal green line shows {\it Lenstool}'s best computation time. When the blue histogram is below the green line is the the point when {\it Lenstool-HPC} brute force approach to lensing beats {\it Lenstool}'s image transport method.  }
  \label{fig:Benchmark}
  \end{figure*}

The $Chi^2$ benchmarks time the full $Chi^2$ computation and its four main stages: The gradient computation of a grid, the source computation based of the constraints, finding images by delensing and checking the grid and computing the $Chi^2$ based on the found images. The benchmark was distributed and scaled over 128 nodes which was our maximum available number of test nodes. In contrast to the core scaling analysis, due to time-constraint on the allotted server time we could not rerun them multiple times to study the standard deviation. 
The results are summarized in Fig. \ref{fig:Benchmark} and more details can be found in the appendix. The two main stages to pay attention to are the gradient computation and the delensing stage.

\subsubsection{Gradient Computation}


The computation of a 5000x5000 lensing map on one Pizdaint P100 GPU takes only 0.93 seconds. At a single node level, compared to a Helvetios node with 36 cores, the single GPU version outperforms {\it Lenstool-HPC}s CPU version by a factor 5 and {\it Lenstool} by a factor 10. Since we upgraded {\it Lenstool}s single map generation to be distributable at a node level, for the common user the P100 version actually outperforms it by a factor 360.

Fig. \ref{fig:Piz_Daint_GPU_Gradient} and Fig. \ref{fig:Helvetios_CPU_Gradient} show the scaling of the gradient computation for CPUs and GPUs up 128 Nodes. Up to 32 nodes the software scales well with a parallelism efficiency of 0.75 . Around 64, for both GPU and CPUs, the scaling worsens with a parallelism efficiency of around 0.5. This is mainly because the shrinking amount of work per node is starting to be insufficient to hide the latencies of the computation. The parallelism efficiency should rise the more complex the problem, but the inverse is also true. The gradient computation could still be distributed over more then 128 nodes for slight gain but we were limited here by the available hardware. 

At 128 nodes, {\it Lenstool-HPC} is 55.3 times faster then it single node GPU version, meaning approximately 500 times faster then {\it Lenstool}'s single map gradient computation. The parallelised CPU version is roughly 4 times slower then its parallelised GPU counter part. It remains competitive enough that even users who do not have access to GPU cluster can generate lensing map efficiently. For cost-conscious users who wish a reasonably high efficiency, with 32 P100 GPUs at one map every 0.04 seconds we can do the Abell\,2744 error computation \citep{2015Jauzac} with 10014 in 166 minutes, a bit less then 3 hours. This is 25 times faster then the {\it Lenstool} version especially tuned for the Benchmarks. 

\subsubsection{Delensing and searching for images}

As stated above, this task is not trivially parallelisable. While the work can distributed over the different GPUs, the divergent execution path that appears when an image is found, impacts the parallelism efficiency quickly. Already at 4 GPU nodes (see Fig. \ref{fig:Piz_Daint_GPU_Delense}), we are at an efficiency of 0.73 and seem to saturate around 8 nodes. At a single node level this does not impact us much (see Fig. \ref{fig:Piz_Daint_GPU_Total}). The task takes only $11\%$ of the total runtime, with the rest going to the gradient computation. However, since the gradient is scaling well, already on 16 cores the delensing task takes $27\%$ of the total runtime with noticable effects. The parallelism efficiency of the total $\chi^2$ computation starts to drop around 8 nodes by the delensing task before it saturates completely around 32 nodes. 
This final saturation is not only due to worsening of the gradient computation efficiency. The computation time of the gradients over 32 to 128 GPUs simply has reached the same level as the computing of the sources on the CPUs around 0.04 to 0.02 seconds. Since the delensing task is depended on both gradient computation and source computation, it cannot start without both having finished running, creating the observed saturation. 

The CPU version in contrast shows a lot less degradation to its parallelism efficiency, at least up to 64 nodes. This corresponds to our expectation since CPUs have less but faster computation units then GPUs. The amount of work per core never reaches a stage when it is insufficient to hide the latencies of the divergent execution paths. CPUs compilers have also been already heavily optimized to handle these complex operations. 

With this in mind, the {\it Lenstool-HPC}'s brute force GPU version manages to beat {\it Lenstool} fully recurrent image transport with only 4 GPUs (see Fig. \ref{fig:Piz_Daint_GPU_Total}) and can still scale up 32 for a total speedup of 5.9. {\it Lenstool-HPC} brute force CPUs version is less successful, managing to beat {\it Lenstool} only with 64 nodes for a speed-up of 1.7 but also demonstrates more parallel effiency. Depending on the hardware developments of the future, they could become a extremely credible option.

\section{Conclusions}

We have shown that it is possible to use modern HPC based programming to greatly speed up conventional gravitational lens mass modeling software. On P100 GPUs et SLK CPUs the new {\it Lenstool-HPC} GPU based library has shown to be 360 times faster than {\it Lenstool} on single map computation and 10 times faster on multi-map computation with only a single GPU. The necessary gradient computation have shown to scale extremely well up to 64 nodes with a Hubble Frontier Fields' size problem, generating an additional corresponding speed-up. The brute force implementation proposed for the mass-model $\chi^2$ computation beats {\it Lenstool} recursive but tricky to use image-tranport implementation with only 4 GPUs and scales reasonably well up to 32 nodes. Additionally {\it Lenstool-HPC} non recursive HPC implementation of lens-modelling tools will scale with future hardware developments, ensuring future speed-ups that recursive options will not have. Future development will go towards the full integration of the library into {\it Lenstool} and optimisation of the last bottle necks, in particular the (MCMC) optimisation process. This will be combined with a thorough comparison to other GPU and non GPU based Lens-modelling tools to assess and further improve Lenstool-HPCs Lens-modeling process. 

The achieved speed-up are key to continue using {\it Lenstool} for clusters with many constraints, and to allow a fast evaluation (through the lensing maps) of the quality and properties of the lensing mass models computed. As an example, having a fast lensing maps computation allows quick evaluation of the lensing model and the identification of where the fit is good or bad, allowing us to focus on the modeling. Ultimately, a fast code will allow to address the "bad RMS" of models (typically larger than 10$\times$ the Hubble image resolution) and understand its origin.

The C++ and CUDA based library is publicly available on Github  \url{https://git-cral.univ-lyon1.fr/lenstool/LENSTOOL-HPC} .

\section{Acknowledgments}
CS thanks Mathilde Jauzac for fruitful discussions on Lens-modelling. CS also acknowledges support from the ESA-NPI grant 4000120530/17/NL/MH, and the SNF Sinergia "Euclid" FNS CRSII5\_173716. GF gratefully acknowledges support from the EPFL Facult\'{e} des Sciences de Base. This work was supported by EPFL through the use of the facilities of its Scientific IT and Application Support Center. The authors gratefully acknowledge the use of facilities of the Swiss National Supercomputing Centre (CSCS), in particular Colin McMurtrie and Hussein Nasser El-Harake for their constant support on the Greina test cluster where most of the GPU development was performed. This research made use of matplotlib \citep{Hunter2007}, Inkscape, TeX Live, and NASA's Astrophysics Data System. 

\section*{References}

\bibliographystyle{model2-names.bst}
\bibliography{Lenstool-HPC_revised}

\appendix
\onecolumn
\section{Benchmark results}\label{appendix_a}

The following tables and figures contain all information of the Benchmarks we did for {\it Lenstool-HPC} run on the clusters summarized in table \ref{table:Hardware}.

  \begin{figure}[h]
    \begin{subfigure}[b]{0.5\textwidth}
        \includegraphics[width=\textwidth]{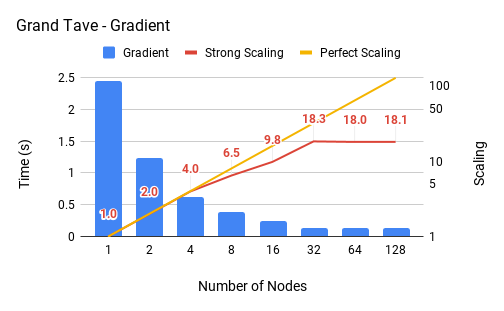}
        \caption{Benchmark results for the deflection gradient computation using the CPU version on the Grand Tave cluster.}
        \label{fig:Grand_Tave_Gradient}
    \end{subfigure}
    \begin{subfigure}[b]{0.5\textwidth}
        \includegraphics[width=\textwidth]{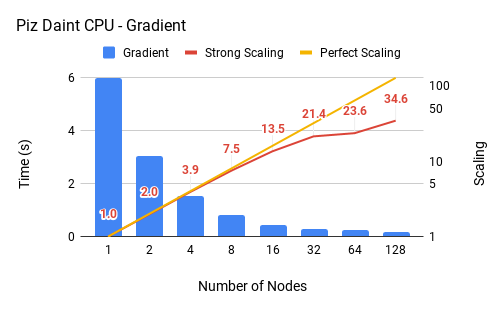}
        \caption{Benchmark results for the deflection gradient computation using the CPU version on the Pizdaint cluster.}
        \label{fig:Piz_Daint_CPU_Gradient}
    \end{subfigure}
    \begin{subfigure}[b]{0.5\textwidth}
        \includegraphics[width=\textwidth]{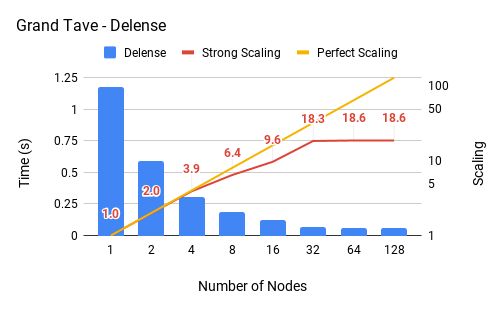}
        \caption{Benchmark results for the constraint delensing operation using the CPU version on the Grand Tave cluster.}
        \label{fig:Grand_Tave_Delense}
    \end{subfigure}
    \begin{subfigure}[b]{0.5\textwidth}
        \includegraphics[width=\textwidth]{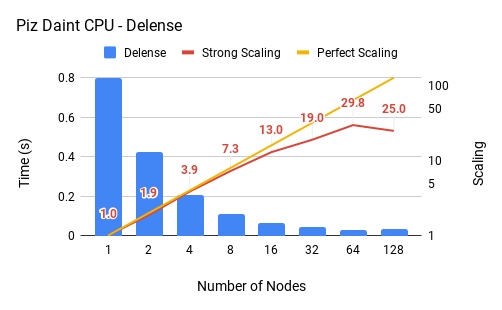}
        \caption{Benchmark results for the constraint delensing operation using the CPU version on the Pizdaint cluster.}
        \label{fig:Piz_Daint_CPU_Delense}
    \end{subfigure}
    \begin{subfigure}[b]{0.5\textwidth}
        \includegraphics[width=\textwidth]{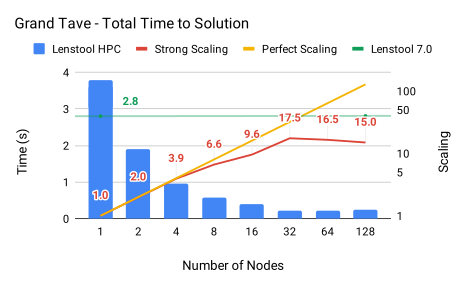}
        \caption{Benchmark results for the total computation time using the hybrid GPU-CPU version on the Grand Tave cluster.}
        \label{fig:Grand_Tave_Total}
    \end{subfigure}
    \begin{subfigure}[b]{0.5\textwidth}
        \includegraphics[width=\textwidth]{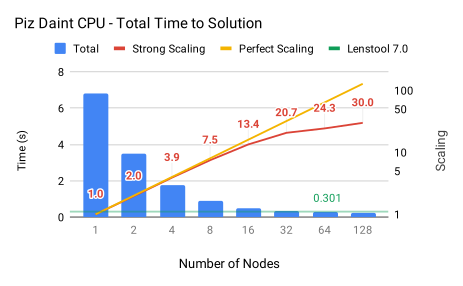}
        \caption{Benchmark results for the total computation time using the hybrid GPU-CPU version on the Pizdaint cluster.}
        \label{fig:Piz_Daint_CPU_Total}
    \end{subfigure}
  \caption{Benchmark and Scaling results of {\it Lenstool-HPC} on Pizdaint CPU (CSCS) and Grand Tave  (CSCS): the blue histogram shows the time results in function of the number of computation units (nodes) used. The yellow and red line indicate respectively the ideal and actual scaling of Lenstol-HPC computation time in function of nodes used. The horizontal green line shows {\it Lenstool}'s best computation time. When the blue histogram is below the green line is the the point when {\it Lenstool-HPC} brute force approach to lensing beats {\it Lenstool}'s image transport method.  }
  \label{fig:Pizdaint}
  \end{figure}

\begin{table}[h]
\caption{Core scaling analysis results on Helvetios on AVX512 capable machines. The results shown are the mean of 10 runs}             
\label{table:Helvetios_Core}  
\setlength{\tabcolsep}{2pt}
\begin{tabular}{l|cc|cccccccc}
\hline
\multicolumn{9}{c}{Core Scaling results Helvetios AVX512}  \\
\hline
& \multicolumn{2}{c}{{\it Lenstool}} &   \multicolumn{6}{c}{{\it Lenstool-HPC}} \\
Cores & {\it Lenstool} [s] & Scaling & AOS [s]& Scaling & SOA novec [s]& Scaling & SOA AVX2 [s] & Scaling & SOA AVX512 [s] & Scaling\\
1 & 281.92 $\pm$ 0.051 & 1.0 & 235.47 $\pm$ 0.040 & 1.0 & 307.80 $\pm$ 0.069 & 1.0 & 107.04 $\pm$ 0.067 & 1.0 & 65.09 $\pm$ 0.068 & 1.0 \\
2 & 146.79 $\pm$ 0.428 & 1.9 & 122.71 $\pm$ 0.539 & 1.9 & 160.21 $\pm$ 0.504 & 1.9 & 55.90 $\pm$ 0.091 & 1.9 & 34.07 $\pm$ 0.051 & 1.9\\
3 & 99.68 $\pm$ 0.155 & 2.8 & 83.43 $\pm$ 0.178 & 2.8 & 108.74 $\pm$ 0.085 & 2.8 & 38.20 $\pm$ 0.027 & 2.8 & 23.35 $\pm$ 0.246 & 2.8\\
4 & 76.04 $\pm$ 0.103 & 3.7 & 63.88 $\pm$ 0.097 & 3.7 & 83.25 $\pm$ 0.055 & 3.7 & 29.23 $\pm$ 0.059 & 3.7 & 19.10 $\pm$ 0.262 & 3.4\\
5 & 62.07 $\pm$ 0.002 & 4.5 & 51.32 $\pm$ 0.001 & 4.6 & 66.85 $\pm$ 0.001 & 4.6 & 23.54 $\pm$ 0.003 & 4.6 & 16.21 $\pm$ 0.020 & 4.0\\
6 & 51.16 $\pm$ 0.003 & 5.5 & 43.16 $\pm$ 0.002 & 5.5 & 55.79 $\pm$ 0.002 & 5.5 & 19.66 $\pm$ 0.004 & 5.4 & 13.56 $\pm$ 0.005 & 4.8\\
7 & 43.86 $\pm$ 0.001 & 6.4 & 36.72 $\pm$ 0.003 & 6.4 & 47.88 $\pm$ 0.002 & 6.4 & 17.03 $\pm$ 0.045 & 6.3 & 11.68 $\pm$ 0.048 & 5.6\\
8 & 38.31 $\pm$ 0.002 & 7.4 & 32.37 $\pm$ 0.002 & 7.3 & 41.83 $\pm$ 0.002 & 7.4 & 15.90 $\pm$ 0.127 & 6.7 & 11.21 $\pm$ 0.059 & 5.8\\
9 & 34.12 $\pm$ 0.006 & 8.3 & 28.60 $\pm$ 0.002 & 8.2 & 37.23 $\pm$ 0.003 & 8.3 & 14.44 $\pm$ 0.001 & 7.4 & 10.48 $\pm$ 0.037 & 6.2\\
10 & 30.81 $\pm$ 0.030 & 9.1 & 25.75 $\pm$ 0.016 & 9.1 & 33.48 $\pm$ 0.010 & 9.2 & 13.76 $\pm$ 0.009 & 7.8 & 9.44 $\pm$ 0.008 & 6.9\\
11 & 28.32 $\pm$ 0.019 & 10.0 & 23.57 $\pm$ 0.032 & 10.0 & 30.62 $\pm$ 0.032 & 10.1 & 11.84 $\pm$ 0.032 & 9.0 & 8.65 $\pm$ 0.019 & 7.5\\
12 & 27.23 $\pm$ 0.153 & 10.4 & 22.50 $\pm$ 0.146 & 10.5 & 29.36 $\pm$ 0.147 & 10.5 & 11.42 $\pm$ 0.072 & 9.4 & 8.59 $\pm$ 0.057 & 7.6\\
13 & 26.13 $\pm$ 0.001 & 10.8 & 21.71 $\pm$ 0.001 & 10.8 & 28.27 $\pm$ 0.001 & 10.9 & 11.13 $\pm$ 0.001 & 9.6 & 8.62 $\pm$ 0.001 & 7.5\\
14 & 24.31 $\pm$ 0.002 & 11.6 & 20.20 $\pm$ 0.005 & 11.7 & 26.40 $\pm$ 0.002 & 11.7 & 10.33 $\pm$ 0.001 & 10.4 & 7.85 $\pm$ 0.002 & 8.3\\
15 & 22.71 $\pm$ 0.004 & 12.4 & 18.88 $\pm$ 0.004 & 12.5 & 24.57 $\pm$ 0.009 & 12.5 & 9.66 $\pm$ 0.004 & 11.1 & 7.20 $\pm$ 0.001 & 9.0\\
16 & 21.64 $\pm$ 0.057 & 13.0 & 17.97 $\pm$ 0.035 & 13.1 & 23.46 $\pm$ 0.028 & 13.1 & 9.18 $\pm$ 0.020 & 11.7 & 6.74 $\pm$ 0.001 & 9.7\\
17 & 20.69 $\pm$ 0.003 & 13.6 & 17.19 $\pm$ 0.000 & 13.7 & 22.46 $\pm$ 0.007 & 13.7 & 8.79 $\pm$ 0.001 & 12.2 & 6.50 $\pm$ 0.002 & 10.0\\
18 & 19.64 $\pm$ 0.012 & 14.4 & 16.25 $\pm$ 0.021 & 14.5 & 21.19 $\pm$ 0.025 & 14.5 & 8.44 $\pm$ 0.001 & 12.7 & 6.15 $\pm$ 0.002 & 10.6\\ 
\end{tabular} 
\end{table}   

\begin{table}[h]
\caption{Core scaling analysis results on FIDIS on AVX2 capable machines. The results shown are the mean of 10 runs}             
\label{table:FIDIS_Core}  
\setlength{\tabcolsep}{8pt}
\begin{tabular}{l|cc|cccccc}
\hline
\multicolumn{9}{c}{Core Scaling results FIDIS AVX2}  \\
\hline
& \multicolumn{2}{c}{{\it Lenstool}} &   \multicolumn{6}{c}{{\it Lenstool-HPC}} \\
Cores & {\it Lenstool} [s] & Scaling & AOS [s]& Scaling & SOA novec [s]& Scaling & SOA AVX2 [s] & Scaling \\
1 & 406.87 $\pm$ 0.003 & 1.0 & 374.36 $\pm$ 0.005 & 1.0 & 515.32 $\pm$ 0.003 & 1.0 & 228.90 $\pm$ 0.001 & 1.0 \\
2 & 203.61 $\pm$ 0.010 & 2.0 & 187.25 $\pm$ 0.002 & 2.0 & 257.95 $\pm$ 0.011 & 2.0 & 115.46 $\pm$ 0.001 & 2.0 \\
3 & 135.92 $\pm$ 0.002 & 3.0 & 124.80 $\pm$ 0.005 & 3.0 & 172.01 $\pm$ 0.012 & 3.0 & 76.38 $\pm$ 0.001 & 3.0 \\
4 & 101.83 $\pm$ 0.008 & 4.0 & 93.58 $\pm$ 0.001 & 4.0 & 129.87 $\pm$ 0.007 & 4.0 & 57.51 $\pm$ 0.000 & 4.0 \\
5 & 81.47 $\pm$ 0.001 & 5.0 & 74.85 $\pm$ 0.002 & 5.0 & 103.23 $\pm$ 0.005 & 5.0 & 45.90 $\pm$ 0.001 & 5.0 \\
6 & 67.96 $\pm$ 0.001 & 6.0 & 62.46 $\pm$ 0.002 & 6.0 & 86.08 $\pm$ 0.003 & 6.0 & 38.36 $\pm$ 0.000 & 6.0 \\
7 & 58.29 $\pm$ 0.001 & 7.0 & 53.57 $\pm$ 0.001 & 7.0 & 73.84 $\pm$ 0.001 & 7.0 & 32.92 $\pm$ 0.001 & 7.0 \\
8 & 51.40 $\pm$ 0.002 & 7.9 & 46.80 $\pm$ 0.001 & 8.0 & 64.98 $\pm$ 0.002 & 7.9 & 29.09 $\pm$ 0.001 & 7.9 \\
9 & 45.34 $\pm$ 0.002 & 9.0 & 41.68 $\pm$ 0.001 & 9.0 & 57.37 $\pm$ 0.001 & 9.0 & 25.81 $\pm$ 0.000 & 8.9 \\
10 & 40.94 $\pm$ 0.001 & 9.9 & 37.47 $\pm$ 0.001 & 10.0 & 51.62 $\pm$ 0.001 & 10.0 & 23.09 $\pm$ 0.000 & 9.9 \\
11 & 37.23 $\pm$ 0.002 & 10.9 & 34.11 $\pm$ 0.001 & 11.0 & 47.24 $\pm$ 0.002 & 10.9 & 21.02 $\pm$ 0.000 & 10.9 \\
12 & 34.16 $\pm$ 0.000 & 11.9 & 31.28 $\pm$ 0.001 & 12.0 & 43.05 $\pm$ 0.001 & 12.0 & 19.26 $\pm$ 0.000 & 11.9 \\
13 & 31.72$ \pm$ 0.001 & 12.8 & 28.87 $\pm$ 0.001 & 13.0 & 39.74 $\pm$ 0.001 & 13.0 & 17.79 $\pm$ 0.000 & 12.9 \\
14 & 29.29 $\pm$ 0.002 & 13.9 & 26.87 $\pm$ 0.001 & 13.9 & 36.97 $\pm$ 0.001 & 13.9 & 16.54 $\pm$ 0.000 & 13.8 \\ 
\end{tabular} 
\end{table}

\begin{table}[]
\caption{Distributed scaling analysis results. The benchmarks were run on the CSCS pizdaint CPU and GPU machines and the EPFL Helvetios and Grand Tave CPU clusters. All results shown are in seconds. }             
\label{table:Distributed_Scaling}  
\setlength{\tabcolsep}{10pt}
\begin{tabular}{llllllllll}
\hline
\multicolumn{9}{c}{Piz Daint@CSCS - GPU}   \\
\hline
Nodes      & Gradient & Source    & Delense     & Comm        & $\chi^2$         & Total        & Strong Scaling & {\it Lenstool} 7.0 &              \\
1                    & 0.93     & 0.0025 & 0.1210    & \SI{1.40e-5}    & \SI{2.1e-5}     & 1.05         & 1.00           & 0.35         &              \\
2                    & 0.47     & 0.0155 & 0.0643    & \SI{7.92e-3}    & \SI{2.3e-5}     & 0.54         & 1.93           & 0.35         &              \\
4                    & 0.24     & 0.0205 & 0.0412    & \SI{7.87e-3}    & \SI{2.5e-5}     & 0.29         & 3.56           & 0.35         &              \\
8                    & 0.12     & 0.0195 & 0.0230    & \SI{8.51e-3}    & \SI{2.8e-5}     & 0.17         & 6.31           & 0.35         &              \\
16                   & 0.06     & 0.0282 & 0.0279    & \SI{8.13e-3}    & \SI{3.2e-5}     & 0.10         & 10.19          & 0.35         &              \\
32                   & 0.04     & 0.0279 & 0.0155    & \SI{8.64e-3}    & \SI{3.2e-5}     & 0.06         & 16.54          & 0.35         &              \\
64                   & 0.03     & 0.0310 & 0.0215    & \SI{9.08e-3}    & \SI{3.3e-5}     & 0.07         & 15.97          & 0.35         &              \\
128                  & 0.02     & 0.0148 & 0.0079    & \SI{9.29e-3}    & \SI{3.6e-5}     & 0.04         & 28.52          & 0.35         &              \\
     
\end{tabular}
\end{table}     
                     
\begin{table}[]
\setlength{\tabcolsep}{10pt}
\begin{tabular}{lllllllll}
\hline
\multicolumn{9}{c}{Piz Daint@CSCS - CPU}    \\
\hline
Nodes         & Gradient & Source    & Delense     & Comm        & $\chi^2$         & Total        & Strong Scaling & {\it Lenstool} 7.0 \\
1                    & 5.99 & 0.0022 & 0.7999    & \SI{2.70e-5}    & \SI{2.1e-5}     & 6.82  & 1.00           & 0.301               \\
2                    & 3.04 & 0.0022 & 0.4247    & \SI{2.06e-4}    & \SI{2.3e-5}     & 3.48  & 1.96           & 0.301               \\
4                    & 1.51 & 0.0030 & 0.2069    & \SI{3.01e-4}    & \SI{2.5e-5}     & 1.75  & 3.90           & 0.301               \\
8                    & 0.80 & 0.0030 & 0.1101    & \SI{1.46e-3}    & \SI{2.8e-5}     & 0.91 & 7.47           & 0.301               \\
16                   & 0.44 & 0.0030 & 0.0616    & \SI{2.14e-3}    & \SI{3.2e-5}     & 0.51 & 13.42          & 0.301               \\
32                   & 0.28 & 0.0030 & 0.0421   & \SI{4.85e-3}     & \SI{3.2e-5}     & 0.33 & 20.67          & 0.301               \\
64                   & 0.25 & 0.0031 & 0.0268   & \SI{8.18e-3}    & \SI{3.3e-5}     & 0.28  & 24.33          & 0.301               \\
128                  & 0.17 & 0.0030 & 0.0320   & \SI{8.55e-3}    & \SI{3.6e-5}     & 0.23 & 29.96          & 0.301               \\
\end{tabular}
\end{table}     
                     
\begin{table}[]
\setlength{\tabcolsep}{10pt}
\begin{tabular}{lllllllll}
\hline
\multicolumn{9}{c}{Helvetios@EPFL}      \\
\hline
Nodes      & Gradient & Source    & Delense     & Comm        & $\chi^2$         & Total        & Strong Scaling & {\it Lenstool} 7.0  \\
1                    & 4.63 & 0.0011 & 0.5894    & \SI{1.50e-5}    & \SI{1.2e-5}     & 5.25  & 1.00           & 0.230               \\
2                    & 2.34 & 0.0017 & 0.2989    & \SI{3.53e-4}    & \SI{1.3e-5}     & 2.65  & 1.98           & 0.230               \\
4                    & 1.18 & 0.0013 & 0.1534    & \SI{4.50e-4}     & \SI{1.1e-5}     & 1.34  & 3.92           & 0.230               \\
8                    & 0.62 & 0.0015 & 0.0787    & \SI{2.47e-4}    & \SI{9.0e-6}     & 0.71 & 7.44           & 0.230               \\
16                   & 0.35 & 0.0015 & 0.0405    & \SI{2.95e-4}    & \SI{9.0e-6}     & 0.39 & 13.42          & 0.230               \\
32                   & 0.20 & 0.0015 & 0.0219    & \SI{8.25e-4}    & \SI{9.0e-6}     & 0.24 & 22.33          & 0.230              \\
64                   & 0.13 & 0.0015 & 0.0135    & \SI{1.28e-3}    & \SI{1.0e-5}      & 0.14 & 37.15          & 0.230         \\
128                  & 0.11 & 0.0015 & 0.0217    & \SI{1.68e-3}    & \SI{1.0e-5}      & 0.13 & 39.15          & 0.230        \\

\end{tabular}
\end{table}     
                     
\begin{table}[]
\setlength{\tabcolsep}{10pt}
\begin{tabular}{lllllllll}
\hline
\multicolumn{9}{c}{Tave@CSCS}      \\
\hline
  Nodes      & Gradient & Source    & Delense     & Comm        & $\chi^2$         & Total  & Strong Scaling & {\it Lenstool} 7.0               \\
1    & 2.45 & 0.0054 & 1.1759    & \SI{1.0e-4}      & \SI{1.73e-4}     & 3.78    & 1.0            & 2.8                        \\
2    & 1.23 & 0.0060 & 0.5921    & \SI{2.1e-4}     & \SI{1.05e-4}     & 1.90    & 2.0            &  2.8                          \\
4    & 0.62 & 0.0061  & 0.3013   & \SI{3.1e-4}    & \SI{1.06e-4}     & 0.96    & 3.9            &  2.8                          \\
8    & 0.38 & 0.0061 & 0.1828    & \SI{1.12e-3}    & \SI{1.08e-4}     & 0.57    & 6.6            &  2.8                          \\
16   & 0.25 & 0.0061 & 0.1219    & \SI{6.33e-3}    & \SI{1.10e-4}      & 0.40    & 9.6            &  2.8                          \\
32   & 0.13 & 0.0062 & 0.0642     & \SI{1.13e-2}    & \SI{1.14e-4}     & 0.22    & 17.5           &  2.8                         \\
64   & 0.14 & 0.0062 & 0.0633    & \SI{2.29e-2}    & \SI{1.14e-4}     & 0.23    & 16.5           &  2.8                          \\
128  & 0.14 & 0.0062 & 0.0632    & \SI{4.67e-2}    & \SI{1.13e-4}     & 0.25    & 15.0           & 2.8                       \\
\end{tabular}
\end{table}

\end{document}